\documentclass[global,twocolumn]{svjour}
\usepackage{amsmath}
\usepackage{latexsym}
\usepackage{graphics}
\usepackage{textcomp}
\usepackage{amstext}
\usepackage{amssymb}
\usepackage{marvosym}
\usepackage{dsfont}
\usepackage{bm}
\usepackage{graphicx}
\usepackage{mathrsfs}

\usepackage{times}

\usepackage[latin1]{inputenc}
\usepackage{fixmath} 
\usepackage[nospace]{cite} 

\renewcommand{\eqref}[1]{(\ref{#1})}
\newcommand{\op}[1]{{\rm #1}}

\newcommand{\bra}[1]{\left\langle#1\right|}
\newcommand{\ket}[1]{\left|#1\right\rangle}
\newcommand{\braket}[2]{\left\langle#1|#2\right\rangle}
\newcommand{\sandwich}[3]{\langle{#1}|{#2}|{#3}\rangle}

\newcommand{\ave}[1]{\left\langle#1\right\rangle}
\newcommand{\dd}{\partial}

\newcommand{\pdiff}[2]{\frac{\dd#1}{\dd#2}}
\newcommand{\diffz}[2]{\frac{\Mdiff^2#1}{\Mdiff#2^2}}
\newcommand{\pdiffz}[2]{\frac{\dd^2#1}{\dd#2^2}}

\newcommand{\calP}{\mathcal{P}}
\newcommand{\calZ}{\mathcal{Z}}

\newcommand{\vek}[1]{\boldsymbol{#1}}

\newcommand{\BS}{\begin{eqnarray}}
\newcommand{\ES}{\end{eqnarray}}
\newcommand{\BSW}{\begin{eqnarray*}}
\newcommand{\ESW}{\end{eqnarray*}}

\newcommand{\indd}[1]{\hspace{-0.6ex}{#1}}

\newcommand{\Mi}{\mathrm{i}} 
\newcommand{\Me}{\mathrm{e}} 
\newcommand{\Mdiff}{\mathrm{d}} 
\newcommand{\Mlabel}[1]{\mathrm{#1}} 

\newcounter{saveeqn}
\newcommand{\alpheqn}{\setcounter{saveeqn}{\value{equation}}\addtocounter{saveeqn}{1}
	\setcounter{equation}{0}%
	\renewcommand{\theequation}{%
		\mbox{\arabic{saveeqn}\alph{equation}}%
		}
	}
\newcommand{\reseteqn}{\setcounter{equation}{\value{saveeqn}}%
	\renewcommand{\theequation}{\arabic{equation}}%
	}

\newcommand{\nlabel}[1]{{\expandafter\edef\csname@currentlabel\endcsname{\arabic{saveeqn}}\csname ltx@label\endcsname{#1}}}

\journalname{Applied Physics B}
\begin{document}

\tolerance=270
\emergencystretch=5pt
\title{Inertial and gravitational mass in quantum mechanics}
\author{E.~Kajari\inst{1}\and
N.~L.~Harshman\inst{1,2}\and
E.~M.~Rasel\inst{3}\and
S.~Stenholm\inst{4,5}\and
G. S\"u{\ss}mann\inst{6}\and
W.~P.~Schleich\inst{1}
}                     
%
%
\institute{
Institut f\"ur Quantenphysik, Universit\"at Ulm, Albert-Einstein-Allee 11, D-89081 Ulm, Germany,
\\ \email{Wolfgang.Schleich@uni-ulm.de}
\and 
Department of Physics, 4400 Massachusetts Ave., NW, American University, Washington, DC 20016-8058, USA
\\ \email{harshman@american.edu}
\and
Institut f\"ur Quantenoptik, Leibniz Universit\"at Hannover, Welfengarten 1, D-30167 Hannover, Germany,
\\ \email{Rasel@iqo.uni-hannover.de}
\and
Physics Department, Royal Institute of Technology, KTH, Stockholm, Sweden,
\and 
Laboratory of Computational Engineering, HUT, Espoo, Finland,
\\ \email{stenholm@atom.kth.se}
\and
Fakult\"at f\"ur Physik der Ludwig-Maximilians-Universit\"at M\"unchen, Schellingstra{\ss}e 4,
D-80799 M\"unchen, Germany, 
\\ \email{Prof.Dr.Suessmann@t-online.de}\\
}
%

%
\maketitle
\begin{abstract}
We show that in complete agreement with classical mechanics, the dynamics of any quantum mechanical wave packet in a linear gravitational potential involves the gravitational and the inertial mass only as their {\it ratio}. In contrast, the spatial modulation of the corresponding energy wave function is determined by the third root of the {\it product} of the two masses. Moreover, the discrete energy spectrum of a particle constrained in its motion by a linear gravitational potential and an infinitely steep wall depends on the inertial as well as the gravitational mass with different fractional powers. This feature might open a new avenue in quantum tests of the universality of free fall.
\end{abstract}


\section{Introduction}
\label{sec:intro}

The equivalence principle is a cornerstone in the foundations of general relativity \cite{MTW}. Indeed, the assumption of the proportionality of inertial and gravitational mass implies that in a linear gravitational potential all bodies experience the same acceleration and fall with the same rate. Without this universality of free fall, the geometrization of gravitation and its reinterpretation as curvature of spacetime would not be possible. The fact that several alternative gravitational theories predict the breakdown of the universality of free fall \cite{Will2001} is one of the main reasons that drives physicists to test this bedrock of modern physics to higher and higher accuracy.

Motivated by the seminal papers on neutron interfero\-metry \cite{Collela1975,Greenberger1979,Bonse1983,Greenberger1983,Rauch2000} and the more recent, impressive matter wave experiments \cite{Kasevich1991,Kasevich1992,Peters1999,Peters2001,Fray2004,Varenna2009,Mueller2010}, we address in the pre\-sent paper the question how the inertial and gravitational mass enter in non-relativistic quantum mechanics. We show that in the case of a linear gravitational potential, quantum dynamics only involves the ratio of the two masses in complete accordance with classical Newtonian mechanics. However, depending on the specific preparation of the initial state, inertial and gravitational mass may appear in a more complicated way in the time evolution of a physical state. As an example of such an initial state, we discuss the energy eigenfunctions in a linear potential \cite{Breit1928,Suessmann1967}, which have been analyzed e.~g. in the context of the coherence of an atom laser \cite{Koehl2001} or in connection with the so-called atom trampoline, also known as the quantum bouncer \cite{Kasevich1990,Aminoff1993,Ovchinnikov1997,Wallis1992,Hughes2009}.
Indeed, the energy eigenstate in this system is non-classical since the corresponding phase space equations for the Wigner function do involve Planck's constant $\hbar$. As a result such states are ideal objects to study the role of inertial and gravitational mass in quantum mechanics. 

Three central results obtained in this paper stand out: (i) The quantum dynamics reduces to classical dynamics and therefore can only involve the ratio of the inertial mass $m_i$ and the (passive) gravitational mass $m_g$, (ii) the spatial modulation of the energy eigenfunctions depends on the third root of the product of the two masses, and (iii) the energy eigenvalues of the gravitational atom trampoline are proportional to $(m_{g}^{2}/m_{i})^{1/3}$.

\subsection{Tests of the universality of free fall}

The universality of free fall, often referred to as ``weak equivalence principle'', states that all bodies experience the same gravitational acceleration independent of their internal structure and composition, provided they are so small in size that one can neglect the effects of gravity gradients. In other words, the (inertial) mass of a body is proportional to its weight, with an universal proportionality constant. A violation of this principle would arise e.~g. when the interaction energy between the nucleons in an atom would not contribute in the same manner to the gravitational mass, as it would for the inertial.

The classical tests \cite{MTW,Will2001} of the universality of free fall assume that the specific gravitational acceleration $\tilde g(A)$ does depend on the internal structure or the composition of the body $A$. This assumption translates in terms of the inertial and gravitational mass into the relation
\begin{equation}
  \tilde g(A)\equiv g \left(\frac{m_g}{m_i}\right)_{\indd{A}}\,.
\label{eq:gtilde}
\end{equation}
Here, the gravitational acceleration $g$ should be considered as a standardized acceleration corresponding to a particular reference body. 

A measure for the breakdown of the universality of free fall is the so-called E\"otv\"os parameter
\begin{equation}
\eta(A,B)\equiv 2\cdot\frac{\left(\frac{m_g}{m_i}\right)_{\indd{A}}-\left(\frac{m_g}{m_i}\right)_{\indd{B}}}{\left(\frac{m_g}{m_i}\right)_{\indd{A}} + \left(\frac{m_g}{m_i}\right)_{\indd{B}}}
=2\cdot\frac{\tilde g(A)-\tilde g(B)}{\tilde g(A)+\tilde g(B)}\,,
\label{eq:Eoetvoes}
\end{equation}
which quantifies the normalized difference in the gravitational accelerations between two different bodies $A$ and $B$.

The first tests of the equivalence of inertial and gravitational mass relied on pendulum experiments and can be traced back to Newton and Bessel \cite{Bessel1832}. A great step towards higher accuracies was realized by the classical torsion balance experiments of E\"otv\"os \cite{Eoetvoes1922} and Roll et al. \cite{Roll1964}. Currently the best upper limits for the E\"otv\"os parameter \cite{Will2005} come from lunar laser ranging on the one hand and from the so-called ''E\"ot-Wash`` experiment \cite{Su1994,Schlamminger2008} on the other. The latter uses a sophisticated rotating torsion balance and limits the E\"otv\"os parameter to 
\begin{equation*}
\eta({\rm Be},{\rm Ti})=(0.3\pm 1.8)\times 10^{-13}
\end{equation*} 
for the gravitational acceleration of Beryllium and Titanium towards Earth. 

The motivation for quantum mechanical tests \cite{Laemmerzahl1996,Viola1997} of the universality of free fall stems from the increase in accuracy that atom interferometry is expected to offer in the future.
Matter wave interferometry with freely falling ${\rm Rb}^{85}$ and ${\rm Rb}^{87}$ isotopes has already been performed \cite{Fray2004}, and several other experiments worldwide using different species of atoms are right now in preparation~\cite{Hogan2009}.

\subsection{Discussion of related work}

In this paper we study the quantum mechanics of a particle in a linear potential. Needless to say, this topic appears prominently in many papers, in particular in connection with the atom trampoline \cite{Kasevich1990,Aminoff1993,Ovchinnikov1997} whose energy eigenstates have been theoretically investigated in \cite{Wallis1992}, as well as in the context of cold neutrons \cite{Nesvizhevsky2002}. 

The work closest to ours is that of Davies \cite{Davies2004} who has investigated the problem of a quantum mechanical particle in a linear gravitational field to gain insight into the equivalence principle. However, motivated by the classical motion his main emphasis is on a thorough analysis of the travel time of wave packets. In complete agreement with our conclusion that the dynamics is classical, he finds that classical and travel times agree far from the classical turning point. However, there are quantum corrections near the turning point. His calculations are based on the definition of the Peres quantum clock~\cite{Peres1980}. 

\subsection{Outline of the paper}

We start in Sect.~\ref{sec:InertialGravitationalMass} by recalling the universality of free fall in Newtonian mechanics. Since we are interested in the Wigner phase space formulation of the corresponding quantum mechanical version, we first introduce in Sect.~\ref{sec:QuantumDynamics} the quantum Liouville equation describing the dynamics of the Wigner function in an arbitrary potential. Moreover, we present the partial differential equations in phase space determining an energy eigenstate in this potential. In Sect.~\ref{sec:UniversalityQM} we then apply these equations to analyze the dynamics of the Wigner function in a linear gravitational potential, as well as to determine the Wigner function of the corresponding energy eigenstates. Whereas the quantum dynamics just reflects the classical time evolution and does not depend Planck's constant $\hbar$, the phase space analog of the energy eigenstates does display quantum features and involves $\hbar$. This fact stands out most clearly in the energy eigenfunctions of the linear potential discussed in Sect.~\ref{sec:Energywavefunction}, where we show that the wave vector governing the spatial modulation of the probability density is determined by the third root of the product the inertial and gravitational mass. In addition, we examine the energy eigenfunctions and the eigenvalues of the atom trampoline, also known as quantum bouncer, that is a particle trapped in the bounded potential resulting from the combination of a linear potential and an infinitely steep wall. We conclude in Sect.~\ref{sec:Discussion} by summarizing our results and by outlining possible experiments.

In order to keep the paper self-contained we summarize concepts pertinent to the present discussion in several appendices. For example in Appendix~\ref{app:Nonpreading} we recall that the time evolution of a particle in a linear potential can be represented in phase space as a product of a shearing operator followed by a displacement. It is only the displacement which contains the gravitational acceleration. This decomposition provides us with deeper insights into the physics of non-spreading Airy wave packets as outlined in Appendix~\ref{App:NonSpreadingAiry}. We dedicate Appendix~\ref{App:Semiclassical} to a discussion of the semi-classical limit of the energy wave function in a linear gravitational potential. Within the Jeffreys--Wentzel--Kramers--Brillouin (JWKB) approximation we regain the universality of free fall. Moreover, we can identify phase space quantization as the origin of the unusual scaling properties of the energy eigenvalues of the atom trampoline in terms of inertial and gravitational mass. We conclude in Appendix~\ref{App:Interferometry} with a phase space analysis of the atomic fountain.


\section{Universality of free fall in Newtonian mechanics}
\label{sec:InertialGravitationalMass}

In the present section we briefly recall Newton's law of motion for a single particle in a linear gravitational potential which shows that the dynamics only depends on the ratio between the gravitational and the inertial mass. For this reason, the free fall of two particles of different compositions is identical provided the inertial and gravitational mass are proportional to each other and the particles start with the same initial conditions. Note that the universality of free fall requires the mentioned proportionality constant to be independent of the particles composition, in which case it can be absorbed in the universal gravitational acceleration. 

Moreover, we examine the dynamics of a classical ensemble of identical particles falling in linear gravitational potential, thereby emphasizing some complications that come along with tests of the universality of free fall based on classical statistical mechanics.

Since we consider a homogeneous gravitational potential, we can restrict our analysis of the dynamics and kinematics to one spatial coordinate $z$. Moreover, instead of dealing separately with the inertial and gravitational mass, we will take advantage of the particle dependent gravitational acceleration~\eqref{eq:gtilde} and denote the inertial mass of a particle most of the time by $m=m_i$.

\subsection{Single particle dynamics}
The time evolution of a particle moving in an external potential $V=V(z)$ follows from Newton's law of motion
\begin{equation}
m_i \,\ddot z=-\pdiff{V}{z}\,,
\label{eq:NewtonEq_of_motion}
\end{equation}
where the dots indicate differentiation with respect to time. 

For a linear gravitational potential 
\begin{equation}
V_l(z)\equiv m_g\,g\, z
\label{eq:LinPotential}
\end{equation} 
we obtain from Eqs.~\eqref{eq:gtilde} and~\eqref{eq:NewtonEq_of_motion}
\begin{equation}
\ddot z = - \frac{m_g}{m_i}\,g\equiv -\tilde g\,.
\label{eq:Eq_of_motion_lin_pot}
\end{equation}
A breakdown of the universality of free fall would manifest itself in a particle dependent gravitational acceleration $\tilde g$. 

The solution of Newton's law of motion~\eqref{eq:Eq_of_motion_lin_pot} reads
\begin{equation*}
 z(t)=z_0+v_0 t -\frac12 \tilde g t^2\,,
\end{equation*}
where $z_0$ and $v_0$ denote the initial position and velocity of the test particle, respectively.
Therefore, tests of the universality of free fall require identical initial conditions for the two test particles.

\subsection{Ensemble dynamics}
\label{sec:EnsembleDynamics}

So far, we have concentrated on the dynamics of a single particle in a linear gravitational field with a well defined initial position $z_0$ and a well defined initial velocity $v_0$. However, in reality it is impossible to prepare the state of the physical system with arbitrary accuracy. For this reason we now consider the dynamics of an ensemble of particles described by a classical distribution function
\begin{equation*}
f_0=f_0(z,v)\,.
\end{equation*}
The probability to find the particle between $z$ and $z+\Mdiff z$ with a velocity between $v$ and $v+\Mdiff v$ is given by $f_0(z,v)\,\Mdiff z\,\Mdiff v$. The probability interpretation requires that $f_0$ is positive everywhere.

Next we turn to the dynamics of the initial ensemble due to a conservative force 
\begin{equation*}
 F(z)=-\pdiff{V}{z}
\end{equation*}
originating from the potential $V=V(z)$. 

The requirement of conservation of probability leads us to the classical Liouville equation
\begin{equation}
\left(\pdiff{}{t}+v\pdiff{}{z}-\frac{1}{m}\pdiff{V}{z}\pdiff{}{v}\right) f(z,v;t)=0
\label{eq:Cl_Liouville}
\end{equation}
subjected to the initial condition $f_0(z,v)\equiv f(z,v;t=0)$.

For the linear potential $V_l$, given by~\eqref{eq:LinPotential}, this equation takes the form
\begin{equation}
\left(\pdiff{}{t}+v\pdiff{}{z}-\tilde g\pdiff{}{v}\right) f(z,v;t)=0
\label{eq:Cl_LiouvilleLin}
\end{equation} 
where we have recalled the definition of specific gravitational acceleration~\eqref{eq:gtilde}.

It is easy to verify, that the solution of~\eqref{eq:Cl_LiouvilleLin} reads
\begin{equation*}
 f(z,v;t)=f_0\Big(z-v t-\frac12\tilde g\,t^2,v+\tilde g t\Big)\,.
\end{equation*}

This expression brings out most clearly the fact, that the dynamics of $f_0$ only depends on $\tilde g$, that is on the ratio of gravitational and inertial mass. However, this property does not exclude the possibility, that $f(z,v;t)$ can contain in addition a dependence on the inertial mass, since the initial distribution $f_0$ might involve the inertial mass.

For example, the stationary solution 
\begin{equation}
f_{\rm s}(z,v)=\mathscr{N}\op{exp}\left[-\frac{m\,v^2}{2 k_{\rm B} T}-\frac{U(z)}{k_{\rm B} T}\right]
\label{eq:Boltzmann}
\end{equation}
of the Boltzmann equation representing a gas of colliding particles at temperature $T$ in a trapping potential $U=U(z)$  involves the inertial mass $m$. Here, $\mathscr{N}$ and $k_{\rm B}$ denote a normalization factor and the Boltzmann constant, respectively. 

When we take $f_{\rm s}$ as the initial distribution of our ensemble of particles propagating in the gravitational field, the final distribution $f(z,v;t)$ will obviously involve not only $\tilde g$ but also the inertial mass $m$. Hence, in a comparison of the free fall of two ensembles of particles of different composition, it is important to ensure that the initial distributions are identical. For the example of two different species of atoms, prepared in the stationary solution of the Boltzmann equation, given by~\eqref{eq:Boltzmann}, this requirement implies that the temperatures and binding potentials have to be adjusted appropriately.

Hence, in the case of two initial ensembles of particles described by a distribution function in position--velocity space, 
a test of this universality of free fall is more complicated. Although each member of the ensemble satisfies the universality of free fall, the two initial distributions have to be identical in order to create two comparable situations.


\section{Wigner function: a few facts}
\label{sec:QuantumDynamics}

In classical mechanics kinematics describes motion without going into the origin of the motion. On the other hand dynamics asks for the origin of the motion. In the same spirit quantum kinematics describes the quantum states and quantum dynamics their time evolution. Throughout our paper, this distinction will be reflected in the separate treatment of initial states and their time evolution. In particular, we consider energy eigenstates as natural candidates for initial states. 

To the best of our knowledge the distinction between kinematics and dynamics has been spelled out for the first time most clearly by Weyl in his book ``The Theory of Groups and Quantum Mechanics'' \cite{Weyl1928}. It is interesting that in this book, Weyl also defines the concept of averages of symmetrically ordered operators using a distribution function, which later became the Wigner function \cite{Dahl1983,Dahl1986,Schleich2001}.
It is this phase space function which we use in our quest to analyze how the inertial and the gravitational mass manifest themselves in quantum mechanics. We devote the present section to a brief review of the Wigner distribution and focus on the elements most pertinent to the present discussion: the quantum Liouville equation and the phase space analog of the time independent Schr\"o\-dinger equation. 

\subsection{Definition}

The Wigner function $W=W(z,p;t)$ is a quasi-probability distribution which lives in phase space spanned by the position $z$ and its conjugate variable, the momentum $p$. When the state of the quantum system is described by a density operator $\hat \rho=\hat \rho(t)$, the corresponding Wigner function reads 
\begin{equation}
 W(z,p;t)\equiv\frac{1}{2\pi\hbar}\int\limits_{-\infty}^\infty\Mdiff\xi\,  \Me^{-\Mi p \xi/\hbar}\,
\sandwich{z+\xi/2\,}{\,\hat \rho(t)}{\,z-\xi/2}\,,
\label{eq:Def_Wignerfkt}
\end{equation}
where $\ket{z}$ denotes a position eigenstate. 

This expression brings out the fact that the Wigner function is real. However, it is not necessarily positive. Moreover, the Wigner function satisfies the marginal properties 
\begin{equation}
\int\limits_{-\infty}^{\infty} \Mdiff p \; W(z,p;t)=\sandwich{z}{\,\hat \rho(t)}{z}\equiv P(z;t)
\label{eq:Wigner_marginalp}
\end{equation}
and
\begin{equation*}
\int\limits_{-\infty}^{\infty} \Mdiff z \; W(z,p;t)=\sandwich{p}{\,\hat \rho(t)}{p}\equiv \tilde P(p;t)\,,
\end{equation*}
that is the integrals over the phase space variables $p$ and $z$ yield the corresponding quantum mechanical probability densities $P=P(z;t)$ and ${\tilde P=\tilde P(p;t)}$ of the conjugate variables.

The definition~\eqref{eq:Def_Wignerfkt} of the Wigner function suggests that this formulation of quantum mechanics rests on the Schr\"odinger representation and requires a wave function or a density operator as a starting point. However, this impression is misleading. The Wigner phase space formulation of quantum mechanics is an approach in its own right. In principle, there is no need to resort to wave functions or density operators. For a more detailed introduction to the Wigner function we refer to \cite{Schleich2001}.

\subsection{Quantum dynamics in phase space}

The dynamics of a quantum state $\ket{\psi(t)}$ describing the motion of a non-relativistic quantum particle of inertial mass $m$ in a potential $V=V(z)$ follows from the Schr\"odinger equation 
\begin{equation}
\Mi \hbar \pdiff{}{t}\ket{\psi(t)}=\hat H\ket{\psi(t)}
\label{eq:Schroedinger_eq}
\end{equation} 
with the Hamiltonian 
$$
\hat H\equiv\frac{\hat p^2}{2 m}+V(\hat z)\,.
$$

This description is equivalent to the quantum Liouville equation 
\begin{equation}
\left(\pdiff{}{t}+\frac{p}{m}\pdiff{}{z}-\pdiff{V}{z}\pdiff{}{p}-\hat{\mathscr{L}}_{\mathrm{o}}\right)
W(z,p;t)=0
\label{eq:QU_Liouville}
\end{equation}
which governs the time evolution of the Wigner function $W=W(z,p;t)$.
Here the differential operator
$$
\hat{\mathscr{L}}_{\mathrm{o}}\equiv\sum_{l=1}^\infty \frac{(-1)^l}{(2l+1)!}
\left(\frac{\hbar}{2}\right)^{\hspace{-0.3ex} 2 l} 
\pdiff{^{2l+1} V(z)}{z^{2l+1}}\pdiff{^{2l+1}}{p^{2l+1}}
$$
involves only odd derivatives of the potential $V$ and even powers of $\hbar$.

Although the quantum Liouville equation is in general derived from the time dependent Schr\"odinger equation~\eqref{eq:Schroedinger_eq}, we could also interpret~\eqref{eq:QU_Liouville} as the equation of motion for the Wigner function without any reference to the Schr\"odinger formulation of quantum mechanics. Indeed, if we possess a priori knowledge about the initial Wigner function, there is no need to refer to the time dependent density operator $\hat \rho(t)$, since the Wigner function $W(z,p;t)$ contains all information of a quantum system.

\subsection{Quantum kinematics in phase space}

Energy eigenstates $\ket{E}$ of the time independent Schr\"odinger equation 
\begin{equation}
\hat H\ket{E}=E\,\ket{E}
\label{eq:TISE}
\end{equation} 
are the elementary building blocks of quantum mechanics. We now briefly motivate the phase space analog of this equation.

Three steps lead to the partial differential equations in Wigner phase space for an energy eigenstate: (i) multiply~\eqref{eq:TISE} by $\bra{E}$, (ii) apply the Weyl--Wigner correspondence \cite{Schleich2001} and (iii) take the real and imaginary part of the resulting equation. In this way we obtain two partial differential equations for the Wigner function $W_{E} = W_{E}(z,p)$ of the energy eigenstate that must be mutually satisfied. 

The imaginary part yields the time independent quantum Liouville equation
\alpheqn
\begin{equation}
\left(\frac{p}{m}\pdiff{}{z}-\pdiff{V}{z}\pdiff{}{p}-\hat{\mathscr{L}}_{\mathrm{o}}\right) W_E=0
\label{eq:PDE_WignerE2}
\end{equation}
which through $\mathscr{L}_{\rm o}$ contains odd derivatives of the potential only. 

Equation~\eqref{eq:PDE_WignerE2} is the quantum Liouville equation~\eqref{eq:QU_Liouville} with a vanishing time derivative. This feature reflects the fact that an energy eigenstate only picks up a phase during its time evolution and the Wigner function is bi-linear in the state. As a result, this phase factor drops out and the Wigner function of an energy eigenstate is time independent. 

From the real part we obtain the analog
\begin{equation}
\left(-\frac{\hbar^2}{8 m}\pdiffz{}{z}+\frac{p^2}{2 m}+V(z)+\hat{\mathscr{L}}_{\mathrm{e}}\right)W_E=E\;W_E
\label{eq:PDE_WignerE1}
\nlabel{eq:PDE_WignerE}
\end{equation}
\reseteqn
of the time independent Schr\"odinger equation which involves only even derivatives of the potential in the differential operator 
\begin{equation*}
\hat{\mathscr{L}}_{\mathrm{e}}\equiv\sum_{l=1}^\infty \frac{(-1)^l}{(2l)!}
\left(\frac{\hbar}{2}\right)^{\hspace{-0.3ex} 2 l} 
\pdiff{^{2l} V(z)}{z^{2l}}\pdiff{^{2l}}{p^{2l}}\,.
\end{equation*}
It is interesting that both, $\hat{\mathscr{L}}_{\mathrm{e}}$ as well as $\hat{\mathscr{L}}_{\mathrm{o}}$, contain only even powers of~$\hbar$.

\subsection{Constraints on the initial Wigner function}

In order to obtain a unique solution of the quantum Liouville equation~\eqref{eq:QU_Liouville}, we must specify the initial Wigner function $W_0=W_0(z,p)$. However, the choice of $W_0$ is a subtle and context dependent enterprise. According to Planck every quantum state must take up in phase space at least an area $2\pi\hbar$. In the language of Wigner functions this condition assumes the form \cite{Schleich2001} 
\begin{equation*}
2\pi\hbar\leq\left[\;\int\limits_{-\infty}^\infty \hspace{-0.7ex}\Mdiff z 
\hspace{-0.7ex}\int\limits_{-\infty}^\infty \hspace{-0.7ex}
\Mdiff p\; W^2_0(z,p)\right]^{-1}\,,
\end{equation*} 
where the equal sign holds for pure states. Here we can interpret the right hand side of this inequality as the effective area of phase space taken up by a quantum state \cite{Sussmann1997}.

But even if we would choose a normalizable function in phase space that satisfies this inequality, it is not clear that it represents a physical Wigner function, since it must be related to a quantum state with positive semi-definite density operator via the Weyl--Wigner correspondence. For a characterization of the set of all phase space functions that represent physical Wigner functions we refer to \cite{Narcowich1986}, or in the special case of Gaussian phase space functions to~\cite{Simon1987}. We emphasize that the energy eigenvalue equations~\eqref{eq:PDE_WignerE} as well as the dynamical equation~\eqref{eq:QU_Liouville} ensure the existence of a valid initial Wigner function and its time evolution.

Since $\hbar$ appears differently in the dynamical and kinematical equations of Wigner phase space, one could imagine, at least mathematically, an extended phase space theory of quantum mechanics in which both Planck's constants differ from each other. For example, a problem where the concept of two different $\hbar$ turned out to be useful \cite{Koike2009} is Kramer's dilemma and the Langer transformation. However, such an extension would probably lead to a physically inconsistent theory. In this sense there is a parallelism between the equivalence of gravitational and inertial mass in general relativity and the identity of Planck's constants of dynamics and kinematics in quantum mechanics.


\section{Universality of free fall in Wigner phase space}
\label{sec:UniversalityQM}

Next we consider the partial differential equations~\eqref{eq:QU_Liouville} and \eqref{eq:PDE_WignerE} determining the Wigner function from phase space for the case of a linear gravitational potential given by~\eqref{eq:LinPotential}. We again analyze quantum dynamics and kinematics separately.

\subsection{Quantum Liouville equation for a linear potential}

For the linear potential~\eqref{eq:LinPotential} the second and all higher derivatives vanish. 
As a result, the operator $\hat{\mathscr{L}}_{\mathrm{o}}$, containing all odd derivatives of the potential are zero and the equation of motion for the Wigner function~\eqref{eq:QU_Liouville} reduces to
\begin{equation}
\left(\pdiff{}{t}+\frac{p}{m}\pdiff{}{z}-m \tilde g\pdiff{}{p}\right)W(z,p;t)=0
\label{eq:QuLiouvilleLin}
\end{equation}
and no longer involves Planck's constant $\hbar$. In fact, the quantum Liouville equation simplifies to the classical Liouville equation~\eqref{eq:Cl_LiouvilleLin} for the linear potential when we recall the relation
\begin{equation}
p=m v
\label{eq:DefVelocity}
\end{equation} 
between the momentum and the velocity of the particle.

Accordingly, the solution of the quantum Liouville equation~\eqref{eq:QuLiouvilleLin} is given by 
\begin{equation}
W(z,m\,v;t)=W_0\Big(z-v\,t-\frac12 \tilde g\,t^2,m\left[v+\tilde g t\right]\Big)\,,
\label{eq:Wignerfunction_Dynamics}
\end{equation}
with $W_{0}(z,p)\equiv W(z,p;t=0)$ being the initial Wigner function of the quantum system. 

The explicit expression~\eqref{eq:Wignerfunction_Dynamics} for the time evolution of the Wigner function emphasizes again the difference between dynamics and kinematics. Since the quantum Liouville equation reduces to the classical Liouville equation in the case of a linear potential, each point in phase space propagates according to classical mechanics while maintaining its ``weight'' as given by the initial Wigner function $W_{0} = W_0(z,p)$. As a result, the \textit{dynamics} only involves the ratio $m_{g}/m_{i}$. However, the initial state may depend on the inertial or gravitational mass in a nontrivial way and therefore the Wigner function $W(z,p;t)$ may not only involve the mass ratio $m_{g}/m_{i}$. Here we find a complete analogy to the classical treatment of Sect.~\ref{sec:EnsembleDynamics}, with one distinct difference: not every initial quasi-probability distribution $W_0(z,p)$ in phase space corresponds to a possible quantum state.

We emphasize that the time evolution of the Wigner function in a linear potential given by~\eqref{eq:Wignerfunction_Dynamics} can be related in a straightforward manner to the free propagation, as discussed in Appendix~\ref{app:Nonpreading}. One important consequence of this consideration is the fact, that the specific gravitational acceleration $\tilde g$ does not influence the spreading of the wave packet, but only its position along the $z$-axis. In other words, the variance $\Delta z^2\equiv\ave{\hat z^2}-\ave{\hat z}^2$ is independent of $\tilde g$. In contrast, the expectation value $\ave{\hat z}$ does depend on $\tilde g$.

Nevertheless, even a classical time evolution can, under appropriate conditions, display non-classical features due to a non-classical initial state. One prominent example is the shrinking \cite{Bialynicki2002,Dahl2004} of a free, radially symmetric wave packet.

\subsection{Wigner function of an energy eigenstate}
\label{sec:EnergyEigenstate}

Next we turn to quantum kinematics and consider as an initial state the energy eigenstate of a quantum particle in the linear gravitational potential given by~\eqref{eq:LinPotential}. The partial differential equations determining the corresponding Wigner function ${W_{E} = W_{E}(z,p)}$ with energy~$E$ follow from Eqs.~\eqref{eq:PDE_WignerE}. In particular,
the time independent quantum Liouville equation~\eqref{eq:PDE_WignerE2} reduces to
\alpheqn
\begin{equation}
\left(\frac{p}{m}\pdiff{}{z}-m \tilde g \pdiff{}{p}\right)W_E(z,p)=0\,,
\label{eq:EigenvalueEq2}
\end{equation}
whereas the eigenvalue equation~\eqref{eq:PDE_WignerE1} for the Wigner function reads
\begin{equation}
\left\{\!\pdiffz{}{z}+\frac{8 m}{\hbar^2}\!\left[E-\left(\frac{p^2}{2 m}
+m \tilde g z\right)\!\right]\!\right\}W_E(z,p)=0.
\label{eq:EigenvalueEq1}
\nlabel{eq:EigenvalueEq}
\end{equation}
\reseteqn

Since the time independent quantum Liouville equation~\eqref{eq:EigenvalueEq2} represents a homogeneous first order partial differential equation, we can apply the method of characteristics and deduce that $W_E$ can depend on the phase space coordinates $z$ and $p$ only via the classical Hamiltonian
\begin{equation}
H_{l}(z,p)\equiv\frac{p^2}{2 m}+m \tilde g z
\label{eq:Hamilton_function}
\end{equation} 
of a particle in a linear potential.

The particular functional dependence of the Wigner function on $H_{l}(z,p)$ is then determined by the eigenvalue equation~\eqref{eq:EigenvalueEq1}. The full solution \cite{Dahl1983,Dahl1986}
\begin{equation}
W_E(z,p)=N_E\cdot\op{Ai}\left[\left(\frac{8}{\hbar^2 m \tilde g^2}\right)^{\!\frac13}\!
\Big(\,H_{l}(z,p)-E\,\Big)\right]
\label{eq:StationaryW_E}
\end{equation}
is given in terms of the Airy function $\op{Ai}=\op{Ai}(y)$, which satisfies the ordinary differential equation \cite{Abramowitz1964}
\begin{equation}
\left(\diffz{}{y}-y\right)\op{Ai}(y)=0\,.
\label{eq:DiffEqAiry}
\end{equation}
Note that the Wigner function $W_E(z,p)$ is not normalizable. As a result the energy $E$ remains a continuous parameter and $N_E$ depends on $E$. 

\begin{figure}[h]
\begin{center}
\includegraphics[width=0.9\columnwidth]{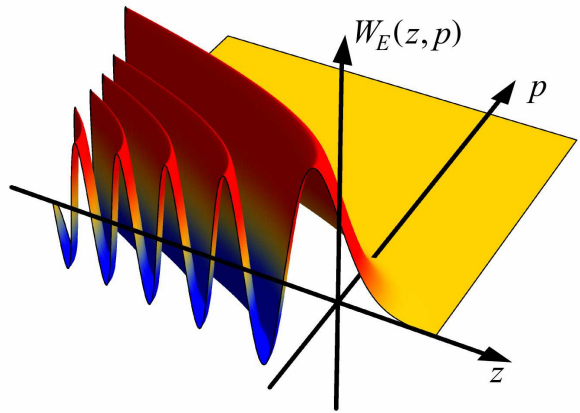}%
\caption{Wigner function $W_E=W_E(z,p)$ of an energy eigenstate in a linear gravitational potential for $E=0$ as given by~\eqref{eq:StationaryW_E}. The parabolic shape is due to the functional dependence of $W_E$ on the Hamiltonian~\eqref{eq:Hamilton_function}. The oscillatory behavior arises from the Airy function which follows from the eigenvalue equation~\eqref{eq:EigenvalueEq1} in phase space.}
\label{fig:Eigenwigner}
 \end{center}
\end{figure}

In Fig.~\ref{fig:Eigenwigner} we depict the Wigner function $W_E=W_E(z,p)$ given by~\eqref{eq:StationaryW_E}.
We recognize a dominant positive-valued ridge along the phase space trajectory $p=\pm p_{\Mlabel{cl}}(z;E)$ given by the classical momentum
\begin{equation}
p_\Mlabel{cl}(z;E)=\sqrt{2 m (E-m \tilde g z)}\,.
\label{eq:p_z_E_classical}
\end{equation} 
following from the condition $H_l(z,p)=E$ with the classical Hamiltonian~\eqref{eq:Hamilton_function}. 
To be precise, at $\pm p_{\Mlabel{cl}}$ the second derivative of the Airy function vanishes.

In the classically forbidden domain of phase space 
\begin{equation*}
E<\frac{p^2}{2 m}+m \tilde g z
\end{equation*} 
which is inaccessible for a classical point particle moving in the linear gravitational potential,
the Wigner function decays exponentially. 

In the classically allowed realm of phase space $W_E$ is oscillatory and can take on negative values. This feature reflects the interference nature of quantum mechanics and expresses the fact that the energy eigenstate is the superposition of a right and a left going wave, as shown in Appendix~\ref{App:Semiclassical}. 

We conclude by noting that the properties of the Airy function have also been studied in the context of tunneling~\cite{Kleber1994}. In particular, the dynamics of the Wigner function tunneling out of a binding delta function potential in the presence of an external static electric field has been studied in~\cite{Czirjak2000}.


\section{Energy wave functions in position space}
\label{sec:Energywavefunction}

The preceding section has employed a phase space analysis to reveal the relationship between quantum and classical behavior of a particle in a linear potential. The dynamics of the Wigner function are entirely classical, whereas the kinematics, and in particular the energy eigenstates, are quantum mechanical. We now address the question how the inertial and gravitational mass make their appearances in the corresponding energy eigenfunctions 
${u_{E} = u_{E}(z)\equiv\braket{z}{E}}$ in {\it position} space. 

In principle, we could obtain the probability density $u^2_{E}(z)$ by taking advantage of the marginal property
of the Wigner function~\eqref{eq:StationaryW_E} together with the integral formula \cite{Berry1977}
\begin{equation*}
\int\limits_{-\infty}^\infty \Mdiff \xi\; \op{Ai}(\xi^2+y)=2^{\frac{2}{3}}\pi\,\op{Ai}^2\big(y/2^{\frac23})\,.
\end{equation*} 

However, it is equally straight forward to solve the corresponding time independent Schr\"odinger equation.
In the present section we pursue this approach and show that the spatial modulation of the energy wave function in the linear gravitational potential depends on the third root of the product of the gravitational and the inertial mass.

\subsection{Unbounded linear potential}
For a particle in a linear potential, the time independent Schr\"odinger equation reads
\begin{equation}
\left(\diffz{}{z}-\frac{2 m}{\hbar^2}\left[m \tilde g\,z-E\right]\right) u_E(z)=0.
\label{eq:SchroedingerEigenval}
\end{equation}
Due to the similarity of this equation with the energy eigenvalue equation in phase space~\eqref{eq:EigenvalueEq1}, their solutions must be similar in form. Indeed, with the help of the differential equation of the Airy function~\eqref{eq:DiffEqAiry}, we can immediately verify that
\begin{equation}
 u_E(z)=\mathscr{N}_E\cdot \op{Ai}\left(\left(\frac{2}{m \hbar^2 \tilde g^2}\right)^{1/3}
\left[m \tilde g z-E\right]\right)
\label{eq:Stationary_u_E}
\end{equation}
satisfies~\eqref{eq:SchroedingerEigenval}. The constant $\mathscr{N}_E$ 
has to be chosen so as to ensure the orthonormality relation 
\begin{equation*}
\braket{E}{E'}=\delta(E-E')
\end{equation*} 
between two different energy eigenstates $\ket{E}$ and $\ket{E'}$.

It is instructive to cast~\eqref{eq:Stationary_u_E} into the form
\begin{equation*}
 u_E(z)=\mathscr{N}_E\cdot \op{Ai}\left(k z-\varepsilon\right)\,,
\end{equation*}
by defining the dimensionless energy
\begin{equation*}
 \varepsilon\equiv E \left(\frac{2}{m\hbar^2 \tilde g^2}\right)^{1/3}
\end{equation*}
and
\begin{equation}
 k\equiv\left(\frac{2 m^2 \tilde g}{\hbar^2}\right)^{1/3}\,,
\label{eq:wavevec}
\end{equation}
which has the same physical units as the familiar wave vector of a plane wave. For $\op{Rb}^{87}$ atoms this quantity defines an inverse length scale of the order of $k\approx 3.3 \times 10^6 \,\op{m}^{-1}$.

When we insert~\eqref{eq:gtilde} into~\eqref{eq:wavevec}, we find that the wave vector
\begin{equation}
 k=\left(\frac{2 \,m_i\,m_g \,g}{\hbar^2}\right)^{1/3}
\label{eq:k_g}
\end{equation}
involves the third root of the product of the inertial and the gravitational mass. 

Therefore, the spatial modulation of the energy eigenstate offers a possibility to compare the masses $m_g$ and $m_i$ by a method independent of the classical experiments based on dynamics. However, it is interesting to note that in the semi-classical limit, there is a revival of the universality of free fall as shown in Appendix \ref{App:Semiclassical}.

We conclude by noting that the energy eigenfunction~\eqref{eq:Stationary_u_E} exhibits a surprising feature when it undergoes a free time evolution. In fact, the free propagation of $u_E$ does not display \cite{Berry1979,Greenberger1980} the phenomenon of spreading, but just an overall acceleration. 
Recently, such Airy wave packets have received great attention in optics and have been realized experimentally with light~\cite{Siviloglou2007a,Siviloglou2007b}. For an explanation of this effect in Wigner phase space we refer to Appendix~\ref{App:NonSpreadingAiry}.

\subsection{Atom trampoline}
\label{sec:AtomTrampoline}

Next we insert an infinite repulsive potential wall at $z=0$. Here, we are not concerned about the nature of this potential wall, i.e.~whether it originates from electromagnetic forces or from gravitational ones that must include the gravitational mass. Its purpose is simply to establish a Dirichlet boundary condition for the wave function and to provide a Hamiltonian that is bounded from below. Moreover, we emphasize that experimental realizations of such a trampoline (or quantum bouncer) for atoms \cite{Kasevich1990,Aminoff1993,Ovchinnikov1997,Wallis1992,Hughes2009}, neutrons \cite{Nesvizhevsky2002}, and light \cite{Valle2009} exist.

The boundary condition
\begin{equation}
u_{E}(0)= 0
\label{eq:Boundary_cond}
\end{equation}
on the wave function~\eqref{eq:Stationary_u_E} enforces the discrete energy eigenvalues
\begin{equation}
E_n=\left(\frac{m \hbar^2 \tilde g^2}{2}\right)^{\frac13}a_{n+1}
\label{eq:Energy_eigenvalues}
\end{equation}
where $n=0,1,2,...$ and $a_j$ denotes the $j$-th zero \cite{Abramowitz1964} of the Airy function, as depicted in Fig. 2.
\begin{figure}[h]
\begin{center}
\includegraphics[width=0.9\columnwidth]{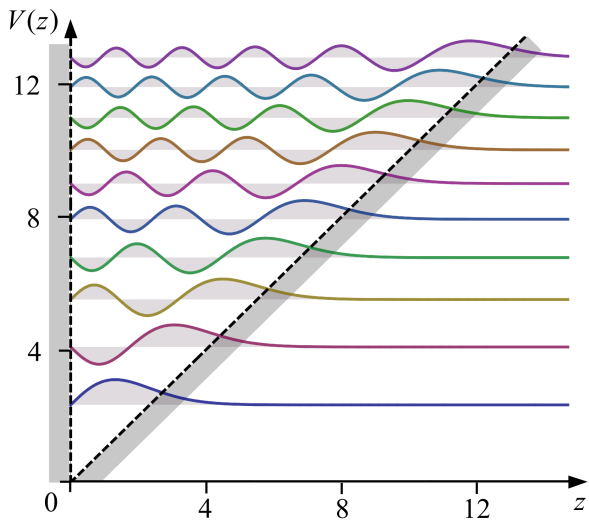}%
\caption{First ten energy eigenfunctions at the corresponding eigenvalues for the atom trampoline. The potential consists of a linear part for $0<z$ and a hard wall at $z=0$.}
 \end{center}
\end{figure}

When we recall the definition~\eqref{eq:gtilde} of $\tilde g$ the energy eigenvalues read
\begin{equation}
E_n=\left(\frac12\,\hbar^2 g^2\right)^{\frac13} m_g^{\frac23}\,m_i^{-\frac13}\,a_{n+1}
\label{eq:Trampoline_E_n}
\end{equation}
and thus depend on $m_{g}^{2/3}$ and $m_{i}^{-1/3}$. 

This is quite a remarkable result because the energy spectrum provides us in principle with a third way to compare the inertial and gravitational mass. However, it is not yet clear which additional degree of freedom of an atom should be coupled to its center-of-mass motion in order to probe the energy spectrum with the necessary energy resolution, since $(m\,\hbar^2\,g^2/2)^{1/3}\approx 2.7\times 10^{-12}\,\op{eV}$ for $\op{Rb}^{87}$ atoms.


\section{Conclusion}
\label{sec:Discussion}

In the present paper we have studied the role of the inertial and gravitational mass in the quantum mechanical treatment of a particle in a linear gravitational potential. Experiments involving the dynamics of wave packets, no matter how complicated the initial state may be, only probe the ratio 
\begin{equation*}
\zeta\equiv\frac{m_{g}}{m_{i}}\,.
\end{equation*} 
This parameter plays a crucial role in the classical experiments on the universality of free fall and is closely connected to the E\"otv\"os parameter~\eqref{eq:Eoetvoes}. 

However, the probability density of a wave packet might well depend on the gravitational and inertial in any arbitrary combination. Using the phase space analysis, we can identify two sources for this fact. (i) The initial Wigner function is a matter of state preparation and thus might involve the inertial mass as well as the gravitational mass. For example, when we start from a particle in a box, the wave function does not depend on the mass at all. In contrast, the energy wave function of the harmonic oscillator depends on square root of the mass. (ii) Due to the marginal property of the Wigner function, the time evolved probability density follows from the integration of the Wigner function and might create in this way new combinations of the gravitational and the inertial mass.

A particularly striking example of this additional freedom is provided by the energy wave function of a quantum particle in linear gravitational potential. Experiments capable of measuring the Airy-function shaped probability density would yield information about $(m_{g} m_{i})^{1/3}$. Moreover, spectroscopy of the discrete energy spectrum of a particle bound in an atom trampoline would provide us with another scaling law $(m_{g}^2/m_{i})^{1/3}$. 

It is interesting to express these scaling laws in terms of $\zeta$ and the geometrical mean 
\begin{equation*}
M^2=m_i\,m_g
\end{equation*}
of the inertial and gravitational mass. Indeed, we find that the spatial modulation of the energy wave function of the linear gravitational potential is sensitive to $M^{2/3}$, whereas the corresponding energy eigenvalues involve the combination $\zeta^{1/2}\,M^{1/3}$.
The additional information concerning the relation between the gravitational and the inertial mass is made possible by quantum mechanics and goes beyond the classical tests of the universality of free fall.

In the past century the spectroscopy of the matter wave representing the electron in the hydrogen atom has triggered the spectacular success of quantum mechanics and quantum Electrodynamics (QED). Indeed, the discrete Balmer series gave birth to matrix mechanics and the Lamb shift led to QED. It would be amusing if in the new century the spectroscopy of matter waves of atoms or Bose--Einstein condensates would shine some new light on the old question of inertial and gravitational mass.


\begin{acknowledgement}

We are grateful to R.~Chiao, R.~F.~O'Connell, J.~P.~Dahl, M.~Efremov, L.~Plimak and M.~O.~Scully for many stimulating and fruitful conversations. We are also grateful to the ECT* for the hospitality provided to us during the workshop {\it ``Many-Body Open Quantum Systems: From Atomic Nuclei to Quantum Dots``}. It was the stimulating atmosphere of this meeting with the spectacular surroundings which allowed us to complete this paper.

As part of the QUANTUS collaboration, this project was supported by the German Space Agency DLR with funds provided by the Federal Ministry of Economics and Technology (BMWi) under grant number DLR 50 WM 0837. One of the authors, E.~Kajari, would like to thank R.~Mack, W.~Fiebelkorn and B.~Casel for help and technical support in the preparation of the manuscript.  Finally, N.L. Harshman acknowledges gratefully the Deutscher Akademischer Austausch Dienst (DAAD) for its support during his stay in Ulm.

\end{acknowledgement}


\appendix

\renewcommand\appendixname{Appendix}

\section{Time evolution in a linear potential}
\label{app:Nonpreading}

In Sect.~\ref{sec:UniversalityQM} we have presented an exact expression for the time evolution of the Wigner function in a linear potential. The main emphasis of this section was the dependence on the inertial and the gravitational mass. We dedicate the present appendix to represent this dynamics as the product of a shearing and a displacement operator acting on the Wigner function in phase space. This analysis is reminiscent \cite{Schrade1997} of the one in a time dependent harmonic oscillator.

We start by recalling~\eqref{eq:Wignerfunction_Dynamics} for the time dependence
\begin{equation}
W(z,p;t)=W_0\Big(z-\frac{p}{m}\,t-\frac12 \tilde g\,t^2\,,\,p+m\,\tilde g\,t\Big)
\label{eq:Wignerfunction_Dynamics2}
\end{equation}
of the Wigner function in terms of the momentum rather than the velocity variable. 

In the absence of gravity, that is for $\tilde g=0$, equation~\eqref{eq:Wignerfunction_Dynamics2} reduces to
\begin{equation}
 W_{f}(z,p;t)=W_0\Big(z-\frac{p}{m} t\,,\,p\Big)
\label{eq:Wignerfunction_freeDynamics}
\end{equation}
and represents the dynamics of a free particle described by the Hamiltonian
\begin{equation}
\hat H_f\equiv\frac{\hat p^2}{2 m}\,.
\label{eq:FreeHamiltonian}
\end{equation} 

The appearance of $p$ in the first argument of the Wigner function~\eqref{eq:Wignerfunction_freeDynamics} is responsible for the familiar shearing effect of the Wigner function. Due to the marginal property~\eqref{eq:Wigner_marginalp}, the shearing in phase space translates into a dispersion of the wave packet in position space. 

When we introduce the displacement operator
\begin{equation}
\hat{\mathscr{D}}(\calZ,\calP)\,\mathcal{F}(z,p)\equiv \mathcal{F}(z-\calZ,p-\calP)
\label{eq:Displacementop}
\end{equation} 
and the time dependent shearing operator
\begin{equation}
\hat{\mathscr{S}}(t)\,\mathcal{F}(z,p)\equiv \mathcal{F}\Big(z-\frac{p}{m}\,t\,,\,p\Big)
\label{eq:Shearingop}
\end{equation} 
which act on any phase space function $\mathcal{F}=\mathcal{F}(z,p)$, we can represent the time evolution of the Wigner function in a linear gravitational potential~\eqref{eq:Wignerfunction_Dynamics2} in the compact form
\begin{equation}
W(z,p;t)=\hat{\mathscr{D}}(\calZ_l(t),\calP_l(t))\,\hat{\mathscr{S}(t)}\,W_0(z,p)\,.
\label{eq:Decomposition}
\end{equation} 
Here we have introduced the time dependent displacement
\begin{equation}
(\calZ_l(t),\calP_l(t))=\left(-\frac12 \,\tilde g\, t^2\,,\,- m\, \tilde g \,t\right)
\label{eq:Shiftvector}
\end{equation}
containing the specific gravitational acceleration $\tilde g$.

We emphasize that the operator $\hat{\mathscr{D}}$ provides a representation of the Lie group of translations in phase space, whereas $\hat{\mathscr{S}}$ corresponds to the Lie group of shear mappings, accordingly. In particular, the order of $\hat{\mathscr{D}}$ and $\hat{\mathscr{S}}$ is important, as reflected by the identity
\begin{equation}
\hat{\mathscr{S}(t)}\,\hat{\mathscr{D}}(\calZ,\calP)=
\hat{\mathscr{D}}\!\left(\calZ+\frac{\calP}{m}t\,,\,\calP\right)\,\hat{\mathscr{S}(t)}\,.
\end{equation}

\begin{figure}[h]
\begin{center}
\includegraphics[width=0.90\columnwidth]{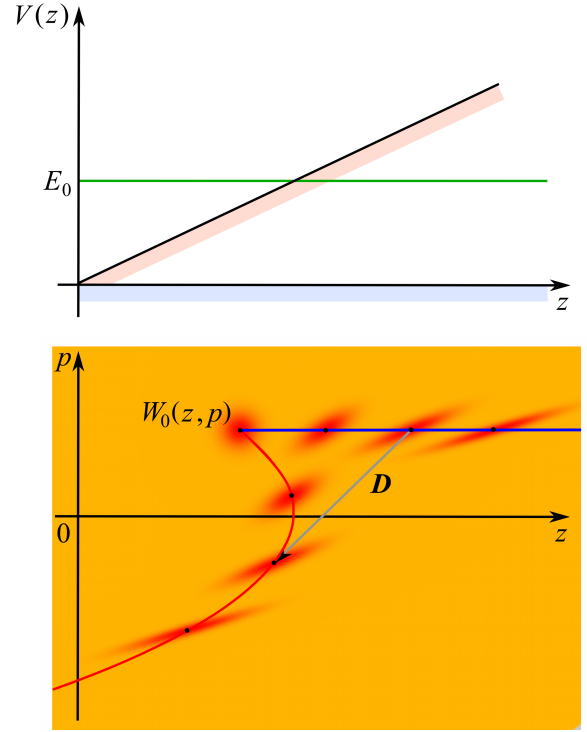}%
\caption{Comparison between the time evolution of a Gaussian Wigner function in phase space (\textit{bottom}) in the presence of a constant or a linear potential (\textit{top}), that is for $V_f(z)\equiv 0$ (\textit{blue}) or $V_l(z)\equiv m\tilde g\, z$ (\textit{red}). The upper picture also indicates the average energy $E_0$ of the wave packet. The center of the Wigner function of the free particle propagates along a straight line (\textit{blue line}), whereas the Wigner function in the linear potential follows a parabola~(\textit{red line}). The time evolved quasi-probability distributions are depicted at three different times and clearly illustrate the fact that the time evolution in a linear potential can be decomposed into a free propagation followed by a shift $\vek{D}=(\calZ_l(t),\calP_l(t))$ in phase space.}
\label{fig:Comparison}
 \end{center}
\end{figure}

In Fig.~\ref{fig:Comparison} we show the time evolution of a Gaussian Wigner function $W_0=W_0(z,p)$ in the presence and absence of a linear potential confirming the decomposition~\eqref{eq:Decomposition} into the product of a time dependent shearing and displacement. An important consequence of this feature is the fact that the linear potential has no influence on the spreading of the wave packet. The spreading is solely due to the free time evolution of the particle given by the Hamiltonian~\eqref{eq:FreeHamiltonian}.

\begin{figure}[h]
\begin{center}
\includegraphics[width=0.9\columnwidth]{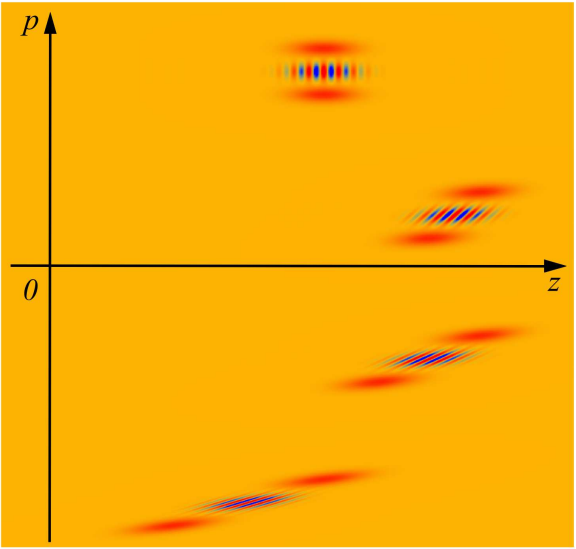}%
\caption{Time evolution in the linear gravitational potential of the Wigner function $W_0(z,p)$ corresponding to a Schr\"odinger cat state consisting of a superposition of two Gaussian states located at the same position but with different initial momenta.}
\label{fig:Springbrunnen}
 \end{center}
\end{figure}

Another surprising implication of~\eqref{eq:Decomposition} is the fact that the linear gravitational potential cannot influence the interference fringes in phase space of a Schr\"odinger cat state that consists of a superposition of two Gaussians centered around the same position but with slightly different initial momenta. In Fig.~\ref{fig:Springbrunnen} we show the time evolution of the Wigner function corresponding to this superposition state. We note that the interference fringes do change in the course of time. However, this change arises solely from the shearing effect and is not due to the gravitational field. Indeed, according to~\eqref{eq:Decomposition} gravity only enters through the displacement of the Schr\"odinger cat obtained from the shearing of free motion.

The property of the Wigner function to track the classical trajectory is not restricted to a time independent linear potential, but can be extended to arbitrary $\tilde g=\tilde g(t)$, resulting in a generalized shift $(\calZ_l(t),\calP_l(t))$ in phase space. We note that this perspective has also been used in the context of many-body theory and the Gross--Pitaevskii equation to separate the center-of-mass motion from the internal dynamics \cite{Kohn1961, Dobson1994, Bialynicki-Birula2002,Nandi2007}.


\section{Non-spreading wave packets}
\label{App:NonSpreadingAiry}

An interesting feature of the energy eigenfunction $u_E$ given by~\eqref{eq:Stationary_u_E} is the fact~\cite{Berry1979,Greenberger1980,Siviloglou2007a,Siviloglou2007b} that it does not spread during the free time evolution governed by the Hamiltonian $\hat H_f$ defined by~\eqref{eq:FreeHamiltonian}. Instead, it preserves its shape and accelerates in positive $z$-direction with a rate $\tilde g\,t^2 /2$.
The representation of the quantum dynamics in a linear potential as a product of shearing and displacement in phase space discussed in Appendix~\ref{app:Nonpreading} offers new insights into the origin of non-spreading Airy-type wave packets.

Indeed, the Wigner phase space description of quantum mechanics allows a rather straight forward derivation of this effect. We elucidate its connection to a specific invariance property of the Hamiltonian~\eqref{eq:Hamilton_function} and show that this symmetry relation defines a broader class of Wigner functions that all correspond to non-spreading wave packets. As an example of this broader class we examine the Wigner function that follows from an incoherent superposition of energy eigenstates~\eqref{eq:Stationary_u_E}. We conclude with an alternative view on the effect of non-spreading wave packets based on the transformation to an accelerated reference frame.

\subsection{Equivalence of shearing and displacement}

The Wigner function~\eqref{eq:StationaryW_E} of an energy eigenfunction is time independent. Thus, by inserting $W_E$ into the time evolution equation~\eqref{eq:Decomposition}, we arrive at the relation
\begin{equation}
W_E=\hat{\mathscr{D}}\,\hat{\mathscr{S}}\,W_E\,.
\label{eq:Decomp_Eigen}
\end{equation} 
When we multiply~\eqref{eq:Decomp_Eigen} by the inverse operator $\hat{\mathscr{D}}^{-1}$ of $\hat{\mathscr{D}}$, we find 
\begin{equation*}
\hat{\mathscr{S}}\,W_E=\hat{\mathscr{D}}^{-1}\,W_E\,.
\end{equation*} 
This equation enjoys an interesting interpretation: the shearing of the Wigner function $W_E$ is equivalent to a displacement of $W_E$ by $(-\calZ_l(t),-\calP_l(t))$. Therefore, $W_E$ preserves its shape during the free time evolution given by~\eqref{eq:Wignerfunction_freeDynamics} and shifts its position in phase space according to
\begin{equation}
 W_f(z,p;t)=W_E\Big(z-\frac12\,\tilde g\, t^2\,,\,p-m\tilde g t\Big)\,.
\label{eq:NonSpreadWigner}
\end{equation}

When we take advantage of the marginal property~\eqref{eq:Wigner_marginalp} of the Wigner function~\eqref{eq:NonSpreadWigner}, we find the identity
\begin{equation}
\int\limits_{-\infty}^\infty \Mdiff p\; W_f(z,p;t)=u^2_E\Big(z-\frac12\,\tilde g\, t^2\Big)\,,
\label{eq:Non_spreading}
\end{equation}
which indicates that the initial probability density of the energy eigenstate~\eqref{eq:Stationary_u_E} does not spread during the free time evolution, but accelerates.

\subsection{Broader class of non-spreading wave packets}

The Wigner function $W_E$ is not the only quasi-probability distribution whose marginal $P(z;t)$ exhibits this interesting feature. Since the non-spreading behavior of the wave packet can be traced back to the invariance relation~\eqref{eq:Decomp_Eigen}, any phase space distribution $W(z,p)$ that obeys the identity
\begin{equation}
W(z,p)=\hat{\mathscr{D}}(\calZ_l(t),\calP_l(t))\,\hat{\mathscr{S}}(t)\,W(z,p)\,.
\label{eq:Symmetry}
\end{equation}
possesses a probability density $P(z;t)$ in position space that does not spread during free time evolution.

The invariance property~\eqref{eq:Symmetry} implies that the corresponding function $W(z,p)$ must depend on the phase space coordinates $z$ and $p$ only via the classical Hamiltonian
\begin{equation}
H_{l}(z,p)\equiv\frac{p^2}{2 m}+m \tilde g\, z
\label{eq:Hamilton_function2}
\end{equation}
of the particle in a linear potential.

Indeed, when we apply the operators $\hat{\mathscr{S}}$ and $\hat{\mathscr{D}}$ to $H_l$, we find with the help of~\eqref{eq:Shiftvector} the relation 
\begin{equation}
\hat{\mathscr{D}}(\calZ_l(t),\calP_l(t))\,\hat{\mathscr{S}}(t)\,H_l(z,p)=H_l(z,p)\,.
\label{eq:HamSym}
\end{equation} 
Hence, any phase space function of the form 
\begin{equation}
W(z,p)=\mathscr{F}(H_l(z,p))
\label{eq:Wignerclass}
\end{equation} 
automatically satisfies the invariance property~\eqref{eq:Symmetry}. 

In order to show that the functions~\eqref{eq:Wignerclass} are the only possible quasi-probability distributions that satisfy the invariance relation~\eqref{eq:Symmetry}, we differentiate~\eqref{eq:Symmetry} with respect to the time $t$ and insert $t=0$. The resulting equation coincides with the time independent quantum Liouville equation~\eqref{eq:EigenvalueEq2}, whose general solution is found by the method of characteristics and exactly coincides with the class of phase space functions~\eqref{eq:Wignerclass}.

We emphasize, that the phase space distributions~\eqref{eq:Wignerclass} cannot be normalized. In fact, when we introduce the new phase space coordinates 
\begin{equation*}
z'\equiv H_l(z,p)\quad\text{and}\quad p'\equiv p\,,
\end{equation*} 
the normalization condition can be expressed as
\begin{equation}
\int\limits_{-\infty}^\infty\!\Mdiff z\!
\int\limits_{-\infty}^\infty\!\Mdiff p\;
W(z,p)=
\frac{1}{m\tilde g}
\int\limits_{-\infty}^\infty\!\Mdiff z' \,\mathscr{F}(z')\,
\int\limits_{-\infty}^\infty\!\Mdiff p'\,.
\end{equation} 
Since the integral over $p'$ diverges, $W$ and therefore its marginals cannot be normalized. 

As a consequence, a wave packet that corresponds to a Wigner function satisfying the invariance relation~\eqref{eq:Symmetry} cannot be exactly realized in an experiment and does not allow the definition of the expectation values $\ave{\hat z}$ and $\ave{\hat p}$. This fact saves the day for Ehrenfest's theorem. 

\subsection{Incoherent superposition of energy eigenstates}

Next, we consider an example for the general class of Wigner functions~\eqref{eq:Wignerclass} with a non-spreading probability density $P=P(z;t)$, namely an incoherent superposition 
\begin{equation}
W_{\op{inc}}(z,p)=\int\limits_{-\infty}^\infty\!\Mdiff E\; g(E)\,W_E(z,p)
\label{eq:IncohSuper}
\end{equation} 
of Wigner functions $W_E$ corresponding to a density operator
\begin{equation}
\hat \rho=\int\limits_{-\infty}^\infty\!\Mdiff E\; g(E)\, \ket{E}\bra{E}\,.
\end{equation} 

For the probability distribution $g=g(E)$ to find the particle in the energy eigenstate $\ket{E}$, we  choose the Gaussian distribution
\begin{equation}
g(E)\equiv \frac{1}{\sqrt{2\pi\sigma^2}}\,\op{exp}\left[-\frac{1}{2\sigma^2}(E-E_0)^2\right]
\label{eq:IncoherentDistr}
\end{equation} 
with mean energy $E_0$ and variance $\sigma^2$.

When we recall~\eqref{eq:StationaryW_E} for the Wigner function $W_E$ together with the 
integral relation
\begin{equation*}
\int\limits_{-\infty}^{\infty}\!\Mdiff\xi \;\frac{e^{-\frac{1}{2\gamma^2}(\xi-\xi_0)^2}}{\sqrt{2\pi\gamma^2}}
\op{Ai}(\xi)=e^{\frac12\gamma^2\left(\xi_0+\frac{\gamma^4}{6}\right)}\op{Ai}\left(\xi_0+\frac{\gamma^4}{4}\right)
\end{equation*} 
for the parameters $\xi_0\equiv\alpha(H_l-E_0)$ and $\gamma\equiv\alpha\sigma$, we obtain the exact formula
\begin{equation}
\begin{split}
W_{\op{inc}}(z,p)=N\cdot e^{\frac12(\alpha\sigma)^2\left[{\alpha\left(\,H_{l}(z,p)-E\,\right)+\frac16(\alpha\sigma)^4}\right]}\\
\times \,\op{Ai}\left[\alpha\left(\,H_{l}(z,p)-E\,\right)+\frac14(\alpha\sigma)^4\right]\,.
\end{split}
\label{eq:IncoherentWigner}
\end{equation} 
Here we have introduced the constant
\begin{equation*}
\alpha\equiv \left(\frac{8}{\hbar^2  m\tilde g^2}\right)^{\!\frac13}\,.
\end{equation*} 

\begin{figure}[h]
\begin{center}
\includegraphics[width=0.9\columnwidth]{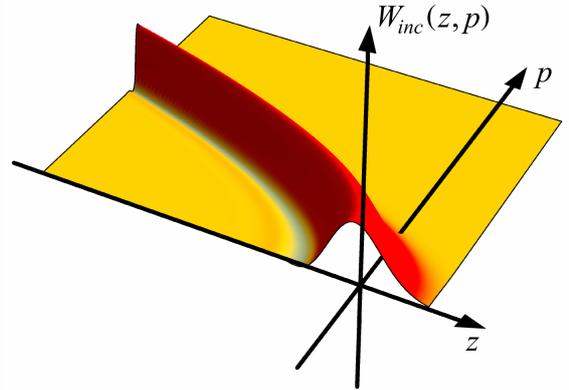}%
\caption{Quasi-probability distribution $W_{\op{inc}}(z,p)$ corresponding to an incoherent superposition of energy eigenstates $\ket{E}$ of a linear potential, that is of Wigner functions~$W_E(z,p)$ with a Gaussian weight function~\eqref{eq:IncoherentDistr} of energy spread $\sigma$ and average energy $E_0$. In contrast to the Wigner function of a single energy eigenstate, this distribution shows strongly suppressed oscillations in the domain $\frac{p^2}{2 m}+m\tilde g z <E_0$ and has a rather broad maximum sightly left to the classical parabolic phase space trajectory corresponding to $E_0$. It is the average over the Gaussian that has eliminated most of the negative contributions in Wigner phase space and has led to a broadening of the maximum of $W_E$.}
\label{fig:EigenwignerIncoherent}
 \end{center}
\end{figure}

The incoherent superposition $W_{\op{inc}}(z,p)$ of a continuous distribution $g=g(E)$ of energy eigenstates $\ket{E}$ given by~\eqref{eq:IncohSuper} differs in general significantly from the Wigner function $W_E(z,p)$ of a single energy eigenstate, as illustrated in Fig.~\ref{fig:EigenwignerIncoherent}. In particular, the domains of phase space where the Wigner function $W_{\op{inc}}$ assumes negative values have almost disappeared. However, due to its functional dependence on $H_l(z,p)$, the parabolic profile has survived.

In the limit $\sigma\rightarrow 0$ for which the energy distribution approaches a $\delta$-function centered around $E_0$, we recover from~\eqref{eq:IncoherentWigner} the Wigner function
\begin{equation*}
\lim\limits_{\sigma\rightarrow 0}W_{\op{inc}}(z,p)=W_{E_0}(z,p)
\end{equation*} 
of the energy eigenstate $\ket{E_0}$.

\subsection{Transformation to an uniformly accelerated frame}

An alternative explanation for the non-spreading wave packet rests on a coordinate transformation~\cite{Greenberger1980} from an inertial to an uniformly accelerated reference frame. It is worthwhile to translate this idea into Wigner phase space in order to obtain yet another interpretation of the invariance relation~\eqref{eq:HamSym}.

The transformation from the inertial coordinate system to the uniformly accelerated reference frame reads
\begin{equation*}
z'\equiv z-\frac12\,\tilde g\, t^2\,,\quad p'\equiv p-m\tilde g t\,,\quad t'\equiv t,
\end{equation*}
with the new phase space coordinates $z'$ and $p'$ and implies the following relations on the partial derivatives
\begin{equation*}
 \pdiff{}{z}=\pdiff{}{z'}\,,\quad \pdiff{}{p}=\pdiff{}{p'}\,,\quad
 \pdiff{}{t}=\pdiff{}{t'}-\tilde g t \pdiff{}{z'} -m\tilde g \pdiff{}{p'}\,.
\end{equation*}
As a result, the quantum Liouville equation 
\begin{equation}
\left(\pdiff{}{t}+\frac{p}{m}\pdiff{}{z}\right)W_f(z,p;t)=0
\label{eq:QLiouvilleFree}
\end{equation}
of the free particle expressed in the accelerated reference frame takes the form
\begin{equation}
 \left(\pdiff{}{t'}+\frac{p'}{m}\pdiff{}{z'}-m \tilde g \pdiff{}{p'}\right)W'(z',p';t')=0\,,
\label{eq:TrafoLiouville}
\end{equation}
with the transformed Wigner function
\begin{equation}
 W'(z',p';t')=W_f\Big(z'+\frac12\,\tilde g\, t'^2\,,\,p'+m\tilde g t';t'\Big)\,.
\label{eq:TrafoWigner}
\end{equation}
The quantum Liouville equation~\eqref{eq:TrafoLiouville} is identical to~\eqref{eq:QuLiouvilleLin} and contains a fictitious  linear gravitational potential arising from the transformation into the accelerated reference frame.

We now assume that our initial quasi-probability distribution for the quantum Liouville equation of a free particle~\eqref{eq:QLiouvilleFree} is given by the Wigner function of an energy eigenstate of a linear potential, that is $W_f(z,p;0)\equiv W_E(z,p)$. The initial Wigner function in the accelerated frame governed by~\eqref{eq:TrafoLiouville} follows from~\eqref{eq:TrafoWigner} for $t'=0$ and takes the form
\begin{equation*}
W'(z',p';0)=W_f(z',p';0)=W_E(z',p')\,,.
\end{equation*}
Since $W_E(z',p')$ satisfies the time independent quantum Liouville equation~\eqref{eq:EigenvalueEq2}, we conclude from~\eqref{eq:TrafoLiouville}
\begin{equation*}
 \pdiff{}{t'} W_E(z',p')=0\,.
\end{equation*}
Hence, the transformed Wigner function exhibits no time evolution and we find
\begin{equation*}
 W'(z',p';t)=W_E(z',p')\,.
\end{equation*}
Insertion of the last expression into~\eqref{eq:TrafoWigner} finally yields
\begin{equation*}
 W_f\Big(z'+\frac12\,\tilde g\, t'^2\,,\,p'+m\tilde g t';t'\Big)=W_E(z',p')\,,
\end{equation*}
which in terms of the original phase space coordinates $z$ and $p$ simply reduces to the Wigner function~\eqref{eq:NonSpreadWigner} of the non-spreading wave packet.

\section{Semi-classical limit of an energy wave function}
\label{App:Semiclassical}

The semi-classical limit of quantum mechanics, that is the JWKB approximation, has always provided deeper insight into the inner workings of quantum theory. In this appendix we apply it to demonstrate that the mass ratio $m_g/m_i$ emerges in the semi-classical limit of the exact energy eigenfunction~\eqref{eq:Stationary_u_E} far from the classical turning point. Moreover, we identify phase space quantization as the origin of the unusual scaling law of the energy eigenvalues of the atom trampoline.

\subsection{Revival of the universality of free fall}

When we recall the asymptotic expansion \cite{Abramowitz1964}
\begin{equation*}
\op{Ai}(-|y|)\cong \frac{1}{\sqrt{\pi}}\frac{1}{\sqrt[4]{|y|}}\cos\left(\frac23|y|^{\frac32}-\frac{\pi}{4}\right)
\end{equation*} 
of the Airy function valid for $1\ll|y|$, we can approximate the energy wave function $u_E=u_E(z)$ given by~\eqref{eq:Stationary_u_E} by a superposition
\begin{equation}
u_E(z)\cong A_E(z)\, \Me^{\Mi \phi(z;E)} + A_E(z)\, \Me^{-\Mi \phi(z;E)}
\label{eq:u_E_superposition}
\end{equation}
of two running waves with identical amplitudes
\begin{equation*}
 A_E(z)=\frac{\mathscr{N}_E}{2\sqrt{\pi}(\varepsilon -k z)^{\frac14}}
\end{equation*}
and opposite phases
\begin{equation}
\phi(z;E)\equiv\frac{1}{\hbar}\int\limits_z^{z_E} \Mdiff \tilde z\; p_\Mlabel{cl}(\tilde z;E)-\frac{\pi}{4}\,.
\label{eq:phasephi}
\end{equation}
Here, $p_\Mlabel{cl}=p_\Mlabel{cl}(z;E)$ denotes the classical momentum~\eqref{eq:p_z_E_classical} and the turning point $z_E$ follows from the condition
\begin{equation*}
 E\equiv m \tilde g z_{E}\,.
\end{equation*}
In this way, the energy wave function $u_E$ can be interpreted as the most elementary matter wave interfero\-meter. Indeed, it consists of a wave running up and one down the linear potential. Both waves have identical amplitudes and their phase difference is governed \cite{Dowling1991} by an area in phase space determined by the strength of the gravitational constant.  However, the representation~\eqref{eq:u_E_superposition} is only justified appropriately away from the turning point, that is for $z\ll z_E$. 

From~\eqref{eq:phasephi} we note the relation
\begin{equation*}
p_\Mlabel{cl}(z;E)=-\hbar \pdiff{\phi}{z}\,.
\end{equation*} 
As a result we can connect the rate of phase change of the JWKB wave given by ~\eqref{eq:u_E_superposition} with the classical momentum $p_\Mlabel{cl}$ at the position $z$.

When we recall the definition~\eqref{eq:p_z_E_classical} of the classical momentum together with the relation~\eqref{eq:DefVelocity} between velocity and momentum, we arrive at the classical velocity 
\begin{equation}
 v_{\op{cl}}=\sqrt{2 \tilde g(z_E-z)}
\label{eq:WKBvelocity}
\end{equation}
of the particle at position $z$, which only involves $\tilde g$, that is the ratio $m_g/m_i$ of the gravitational and the inertial mass. Thus, in the semi-classical limit we obtain a form of the universality of free fall expressed by the relation
\begin{equation*}
 v_{\op{cl}}(z)=-\frac{\hbar}{m}\pdiff{\phi}{z}\,.
\end{equation*}

A position-dependent velocity or momentum does not exist in the Schr\"odinger formulation of quantum  mechanics. Indeed, we either live in position space or in momentum space. However, in the semi-classical limit we can obtain mixed position-momentum variables. It is interesting that this result also follows directly from~\eqref{eq:StationaryW_E} of the Wigner function, as discussed in Sect.~\ref{sec:EnergyEigenstate}.

However, we emphasize that in the neighborhood of the turning point the decomposition~\eqref{eq:u_E_superposition} does not hold true. As a consequence, there the wave function is not able to reproduce~\eqref{eq:WKBvelocity}, where only the ratio of the two masses enters. Around the turning point, the energy wave function thus depends on the third root of the product of the inertial and the gravitational mass.

Our result complements the work of Davies \cite{Davies2004}. Here, the travel time of a wave was calculated based on the Wigner time
\begin{equation}\label{Eq:transittime}
\tau=\hbar \pdiff{\phi}{E}=\int\limits_z^{z_E}\!\Mdiff\tilde z\;\frac{m}{p_\Mlabel{cl}(\tilde z;E)}
=\int\limits_z^{z_E}\!\Mdiff\tilde z\;\frac{1}{v_\Mlabel{cl}(\tilde z;E)}
\end{equation} 
which is valid appropriately away from the turning point of the motion. Indeed, the universality of free fall holds true again and the travel time only involves the ratio of the two masses.

\subsection{Phase space quantization as origin of the scaling law}

The form of the eigenvalues in~\eqref{eq:Energy_eigenvalues} and, in particular, the scaling properties of the two masses in~\eqref{eq:Trampoline_E_n} can also be derived in a straightforward manner from the Kramers
improved Bohr--Sommerfeld rule
\begin{equation}
J\equiv\oint\Mdiff z \;p_\Mlabel{cl}(z;E)=2\pi\hbar\left(n+\frac34\right).
\label{eq:KBSrule}
\end{equation}
The Maslov index is $3/4$ because there is one ``hard'' reflection at $z=0$ contributing 1/2, or equivalently $\pi$ to the phase, and one soft reflection at the classical turning point contributing 1/4, or $\pi/2$ to the phase~\cite{Geldart1986}. 

With the help of the classical momentum~\eqref{eq:p_z_E_classical}, we can evaluate the action
\begin{equation*}
J=\frac{4}{3}\sqrt{\frac{2 }{m \tilde g^2}}\; E^{\frac{3}{2}}
\end{equation*}
which yields, with the quantization condition~\eqref{eq:KBSrule}, the approximate expression
\begin{equation*}
E_n\approx \left(\frac{ m \hbar^2\tilde g^2}{2}\right)^{\frac{1}{3}}
\left[\frac{3\pi}{2}\left(n+\frac34\right)\right]^{\frac{2}{3}}
\end{equation*}
for the energy eigenvalues. Thus, we find the same prefactor as in~\eqref{eq:Energy_eigenvalues} and the approximation \cite{Abramowitz1964}
\begin{equation*}
a_j\cong \left[\frac{3\pi}{8}\left(4j-1\right)\right]^{\frac{2}{3}}
\end{equation*} 
for the $j$-th zero of the Airy function.

This treatment clearly identifies the origin of the unusual scaling laws $m_{i}^{-1/3}$ and $m_{g}^{2/3}$ of the energy eigenvalues: The quantization of the energy levels follows from the quantization of the action, i.e.~of an area in phase space. This quantity involves both masses.

\section{Atomic fountain}
\label{App:Interferometry}

In Appendix~\ref{app:Nonpreading} we have shown that a constant gravitational field has no influence on the phase space interference fringes of a Schr\"odinger cat moving in a linear potential. On the other hand atomic fountains provide us with precision measurements of the gravitational acceleration. In order to bring out the similarities and differences between these two measurement schemes, we first summarize the essential ideas that form the basis of an atomic fountain. In the spirit of the present paper, which relies almost exclusively on the Wigner functions, we then outline the physics of an atomic fountain in phase space, thereby sketching the ideas only. In fact, we do not intend to develop a complete description of this precision instrument. For this purpose we refer e.~g. to~\cite{Borde89,Borde02,Canuel06,Cronin09} and references therein.

\subsection{Basic idea}

In the atomic fountain experiments \cite{Kasevich1991,Kasevich1992}, an effective two-level atom is moving vertically up against the gravitational field of the Earth. Initially, the atom is in its ground state~$\ket{g}$. However, on its way up a laser pulse prepares a coherent superposition of its internal levels. The wave vector $k$ of the laser is aligned along the $z$-axis. Since the transition to the excited state $\ket{e}$ is accompanied by a momentum transfer due to the photon recoil, the atom in the excited state has a different momentum than the atom in the ground state. As a consequence they accumulate different phases during their propagation in the gravitational field. After a time $\tau$ a $\pi$-pulse exchanges the population of the ground and excited state. Finally after another time $\tau$, a third laser pulse mixes the internal levels of the atom and in this way erases the {\it which-way information}. The quantity measured at the end is the probability to find the atom in the ground or excited state.

The measurement scheme used in the atomic fountain is reminiscent of the problem of wave packet interferometry involving two different molecular surfaces~\cite{Garraway1992}. Whereas in typical molecules these potentials are rather complicated, in the atomic fountain they are linear in lowest order. For this reason, the latter can be treated fully analytically.

\subsection{State vector description}

It is straightforward to translate the atomic fountain experiment into the language of state vectors. 
We start with the initial state
\begin{equation*}
\ket{\Psi_i}=\ket{\psi}\ket{g},
\end{equation*} 
where $\ket{\psi}$ represents the center-of-mass motion along the $z$-axis. After the first laser pulse the atom is in the entangled state
\begin{equation}
\ket{\Psi_1}\equiv \frac{1}{\sqrt{2}}\left[\ket{\psi}\ket{g}-i\hat D\ket{\psi}\ket{e}\right]\,.
\label{eq:FirstPulse}
\end{equation}
Here, the unitary operator
\begin{equation}
\hat{D}=e^{i k \hat{z}}
\end{equation} 
imparts the extra momentum from the photon recoil and acts as a displacement of the corresponding Wigner function in phase space due the interaction between atom and laser. We note that we refrain from including additional phases that are imprinted on the atom by the laser which must be taken into account in a real experiment. In~\eqref{eq:FirstPulse}, we also assume that the interaction between the atom and the laser does not otherwise affect the initial center-of-mass wave function $\ket{\psi}$, which provides a good first order approximation for the real situation encountered in atom interferometer experiments.

The time evolution of the atom in the linear gravitational field for the time $\tau$ described by the unitary operator ${\hat U_l\equiv\op{exp}(-i\hat H_l \tau/\hbar)}$ yields the state 
\begin{equation*}
\ket{\Psi_l}\equiv \frac{1}{\sqrt{2}}\left[\hat U_l\ket{\psi}\ket{g}-i\hat U_l\,\hat D\ket{\psi}\ket{e}\right]\,.
\end{equation*} 
The second laser pulse at the time $\tau$ is a $\pi$-pulse and interchanges the probability amplitudes of the ground and excited state which leads to the expression
\begin{equation*}
\ket{\Psi_2}= \frac{1}{\sqrt{2}}\left[-i\hat D\,\hat U_l\ket{\psi}\ket{e}-\hat D^{-1}\hat U_l\hat D\ket{\psi}\ket{g}\right]\,.
\end{equation*}
After another period $\tau$ of unitary evolution, a third laser pulse mixes the internal states with a second $\pi/2$-pulse, and we arrive at the final state
\begin{equation*}
\ket{\Psi_f}=\frac{1}{\sqrt{2}}\Big[\ket{\psi_g}\ket{g}+\ket{\psi_e}\ket{e}\Big]
\end{equation*}
where we have introduced the states
\begin{equation}
\label{eq:groundstate}
\ket{\psi_g}\equiv -\frac{1}{\sqrt{2}}\left[\hat D^{-1}\,\hat U_l\,\hat D\, \hat U_l+\hat U_l\,\hat D^{-1}\,\hat U_l\,\hat D\right]\ket{\psi}
\end{equation}
and
 \begin{equation*}
\ket{\psi_e}\equiv -\frac{1}{\sqrt{2}}\,( i\hat D )\left[\hat D^{-1}\,\hat U_l\,\hat D\,\hat U_l- \hat U_l\,\hat D^{-1}\,\hat U_l\,\hat D \right]\ket{\psi}
\end{equation*}
of the center-of-mass motion of the atom in the ground and excited state, respectively.

\subsection{Determination of the phase difference}

Using the definitions of the momentum displacement operator $\hat{D}$ and the time evolution operator $\hat{U}_l$, the two terms in the expression for the motional states of the atom $\ket{\psi_g}$ or $\ket{\psi_e}$  can be combined and the nature of the interference of the trajectories can be made explicit.  With the notation $\hat D^{-1}\,\hat U_l\,\hat D = \hat U'_l$, equation~(\ref{eq:groundstate}) can be rewritten in the simple form 
\begin{equation*}
\ket{\psi_g}\equiv -\frac{1}{\sqrt{2}}\left[\hat U'_l\, \hat U_l+\hat U_l \,\hat U'_l \right]\ket{\psi}\,.
\end{equation*}
As shown below, using the commutation relation between the position and momentum operator together with the Baker--Campbell--Hausdorff formula, we can prove that
\begin{equation*}
\ket{\psi_g}\equiv -\frac{1}{\sqrt{2}}\left[e^{i k \tilde{g}\tau^2 }+1\right]\,\hat U_l\, \hat U'_l \ket{\psi}\,,
\end{equation*}
which implies the well-known gravity-dependent expression for the probability
\begin{equation*}
P_g=\op{Tr}\left\{\ket{\Psi_f}\bra{\Psi_f}\cdot\ket{g}\bra{g}\right\}=\frac{1}{2}\left(1+\cos\Delta\phi\right)
\end{equation*} 
to find the atom in the ground state. Here, we introduced the phase difference $\Delta \phi\equiv k \tilde{g} \tau^2$.

In order to establish this result, we note that from the expansion of $\hat{U}_l$, one can prove
\begin{equation*}
\hat U'_l =\exp(-i \tau \hat D^{-1}\,\hat H_l\,\hat D/\hbar)\equiv\exp(-i \tau \hat H'_l/\hbar)\,,
\end{equation*}
where the boosted Hamiltonian $\hat H'_l$ is just the original Hamiltonian $\hat H_l$ with the momentum $\hat{p}$ replaced by $\hat{p}+\hbar k$, that is
\begin{equation*}
\hat H'_l= D^{-1}\,\hat H_l\,\hat D = \frac{(\hat{p}+\hbar k)^2}{2 m} + m\tilde{g}\hat{z}.
\end{equation*}
Finally, we use the Baker--Campbell--Hausdorff formula to commute $\hat U'_l$ and $ \hat U_l$ and find
\begin{equation*}
\hat U'_l\, \hat U_l =\hat U_l \,\hat U'_l\,e^{(-i\tau/\hbar)^2[\hat{H}'_l,\hat{H}_l]}=\hat U_l \,\hat U'_l\, e^{ik \tilde{g} \tau^2 }\,,
\end{equation*}
which concludes our simple proof of the phase difference between the two interfering trajectories.

\subsection{Phase space description}

\begin{figure}[h]
\begin{center}
\includegraphics[width=0.9\columnwidth]{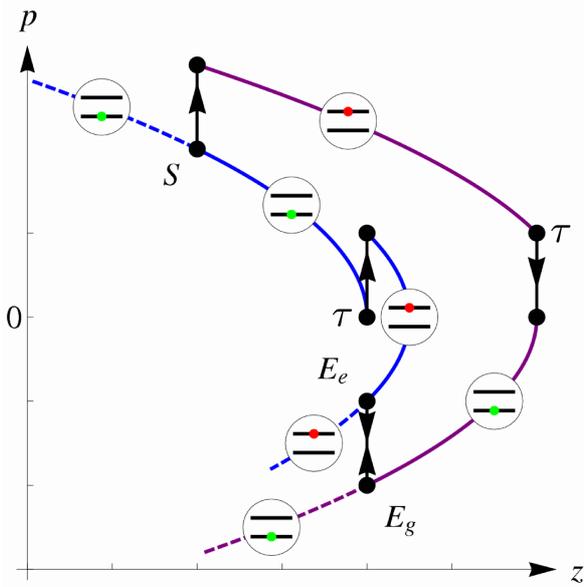}%
\caption{Phase space representation of an atomic fountain experiment. Two $\pi/2$-pulses separated by the time $2\tau$ surround a $\pi$-pulse at time $\tau$ and cause transitions in the internal states of the atom which are accompanied by displacements in phase space along the momentum axis. Each parabola section corresponds to the center-of-mass motion in the linear gravitational field within the propagation time $\tau$. The atom starts its journey in the ground state at the point $S$. Atoms eventually measured in the ground state end up at phase space point $E_g$, whereas atoms detected in the excited state are to be found at the point $E_e$. In both cases two distinct paths lead from the starting point S in phase space to the same final point. The area enclosed by the two paths expressed in units of $\hbar$ turns out to be twice the phase difference $\Delta\phi$ between the two corresponding probability amplitudes.}
\label{fig:GeneralTwoPaths}
 \end{center}
\end{figure}

In Fig.~\ref{fig:GeneralTwoPaths} we represent in phase space the path of the quantum state $\ket{\psi}$ of the center-of-mass motion in the course of time. For an atom that is detected in the ground state $\left|\psi_g\right\rangle$, we have two paths which start from the same point $S$ in phase space and end up at the final point $E_g$. One path corresponds to a sequence of unitary evolution, displacement, unitary evolution and inverse displacement. Therefore, the atom first moves the linear gravitational potential while being in the ground state and then after a transition into the excited state continues its motion in the potential. It concludes with a transition into the ground state at the point $E_g$. The second path starts with the displacement, is followed by unitary evolution and negative displacement, and concludes by the inverse displacement. Here the atom begins its trip by first making a transition into the excited state to be followed by motion in the potential. After de-excitation the atom in the ground state completes its path in the gravitational field. Both paths define an enclosed area in phase space that determines the phase difference between the two amplitudes contributing to $\left|\psi_g\right\rangle$.

Atoms that exit the interferometer in the excited state $\left|\psi_e\right\rangle$ have traversed the same path in phase space as atoms in the ground state except that their final point $E_e$ on the trajectory is different. They either move in the potential, become excited, and then follow again the parabola or become initially excited, follow the parabola, emit, follow again the linear potential, and finally get re-excited.

\begin{figure}[h]
\begin{center}
\includegraphics[width=0.9\columnwidth]{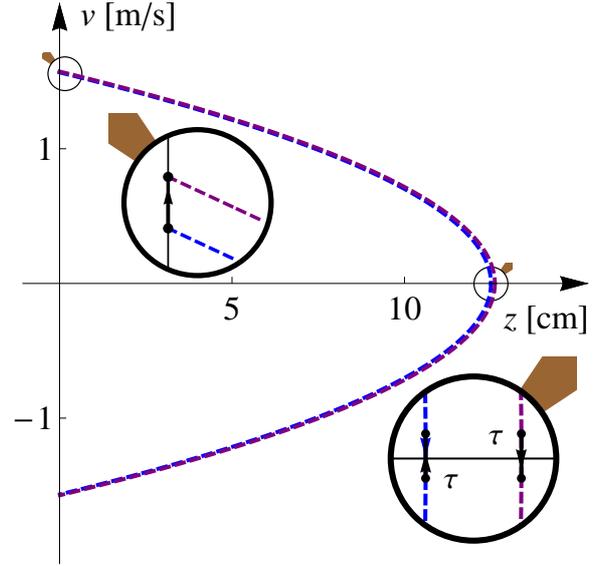}%
{\vspace{1em}\caption{Atomic fountain experiment represented in position-velocity space for typical experimental parameters. Here the momentum transfer due to the absorption or emission of a single photon is much smaller than the typical momentum associated with the motion in the linear gravitational potential. As a result the rather complicated closed curve in phase space shown in Fig.~\ref{fig:GeneralTwoPaths} simplifies more or less to two parabolas (\textit{dotted curves}) which are connected at $z=0$ by vertical lines due to the interaction with the laser (\textit{solid line in upper inset}). At the time $\tau$, the second laser pulse causes another internal transition and a shift in the momentum. Consequently the parabolas are interrupted by straight vertical lines representing the momentum transfer (\textit{lower inset}). For each trajectory we have marked with the symbol $\tau$ the points in phase space that represent the end of the first step of the unitary time evolution.  Note that the straight line on the inner trajectory (\textit{blue dashed line}) is therefore traversed thrice, whereas the corresponding section on the outer path (\textit{purple dashed line}) is traversed only once. The atom initially in the ground state follows the parabola to the point in phase space indicated by $\tau$. The laser pulse moves the atom vertically up in phase space and the excited atom traverses the same vertical line again, but now as part of the unitary time evolution in the potential. This picture is an approximation; in reality the parabolas have a  curvature that for parameters of the experiment is negligible on the length scale of the momentum transfer and invisible even on the magnified scale of the inset.}
\label{fig:SpecificTwoPaths}}
\end{center}
\end{figure}

In Fig.~\ref{fig:SpecificTwoPaths} we show the trajectory of the atom in phase space during a fountain experiment for realistic parameters \cite{Peters2001}. For this case the vertical straight paths due to the momentum exchange with the laser pulse are much shorter than the parabolas corresponding to the classical motion in a linear gravitational potential. 

As a result we can approximate the closed circuit in phase space by two parabolas $p_{\rm cl}(z;E)$ and $p_{\rm cl}(z;E+\Delta E)$ given by~\eqref{eq:p_z_E_classical}, which are connected by two vertical lines of length ${\Delta p=\hbar k}$ at $z=0$. The difference $\Delta E$ in energy is 
to first order in $\Delta p$
\begin{equation}\label{Eq:energydiff}
\Delta E=\frac{p_E}{m}\Delta p =\frac{p_E}{m}\hbar k\,,
\end{equation}
where $p_E=p_{\rm cl}(0;E)$ denotes the momentum at zero potential.

The phase difference $\Delta \phi$ between the two paths turns out to be half of the phase space area enclosed by the two parabolas (expressed in units of $\hbar$).
In order to provide a simple derivation of this result, we first emphasize with the help of~\eqref{eq:phasephi} that $2 \hbar \,\phi(0;E)$ represents the phase space area inside the parabola $p_{\rm cl}(z;E)$ ranging from $z=0$ to the turning point $z_E$.
We then denote by $\delta \phi$ the area of the circuit expressed in units of $\hbar$ and find for small $\Delta E$ the relation
\begin{equation}
\delta \phi\equiv 2\left[\phi(0; E+\Delta E) - \phi(0;E)\right]\approx 2\frac{\partial \phi}{\partial E}\Delta E\;.
\label{eq:DeltaJ}
\end{equation}

When we recall the definition~(\ref{Eq:transittime}) of the classical transit time $\tau$ from $z=0$ to the turning point $z_E$, we find
\begin{equation}
\delta \phi=2\, \frac{\tau}{\hbar}\Delta E\,.
\label{eq:partialJ}
\end{equation}
Finally, substituting Eqs.~(\ref{Eq:energydiff}) into~(\ref{eq:partialJ}) together with the identity $p_E=m\tilde{g}\tau$,  reduces to 
\begin{equation*}
\delta \phi = 2 k\tilde{g}\tau^2=2  \Delta \phi
\end{equation*}
and proves our conjecture in the limit of small momentum transfer $\Delta p\ll p_E$. 
We note that a more thorough analysis shows that this result is actually exact.

This calculation brings to light a remarkable interpretation of the atomic fountain experiment. The phase difference $\Delta \phi$ between the two paths is half the difference $\delta \phi$ between the phases of two energy wave functions in a linear potential of slightly different energies. Each wave represents an interferometer---it is the interference of two counter-propagating waves as expressed by~(\ref{eq:u_E_superposition}). The phase $2 \phi(0;E)$ is the area in phase space enclosed by a straight line at $z=0$ and the parabola corresponding to the energy $E$. The area of the ``crescent moon'' is thus the difference of the areas representing the phases of the WKB energy wave functions in a linear potential.

\subsection{Comparison between cat and fountain}

We conclude by comparing the interference contained in a Schr\"odinger cat such as the one shown in Fig.~\ref{fig:Springbrunnen} and the one in the superposition state in the atomic fountain, depicted in Fig.~\ref{fig:GeneralTwoPaths}. In both cases we consider the quantum state of the center-of-mass motion in a linear gravitational field. However, in the case of the Schr\"odinger cat state, we face the superposition of two states that are located at different points in phase space. This fact stands out most clearly in the language of coherent states $\ket{z+ip}$ of the harmonic oscillator. Indeed, the state
\begin{equation*}
\ket{\psi_{\rm Cat}}=\mathscr{N}_{\rm Cat}\Big[\ket{z+i\,p}+\ket{z+i\,p'}\Big]
\end{equation*} 
of the Schr\"odinger cat corresponds to two Gaussians which differ in their momenta $p$ and $p'$. Here $\mathscr{N}_{\rm Cat}$ denotes a normalization constant. 

In contrast, in the atomic fountain we have the states $\ket{\psi_g}$ and $\ket{\psi_e}$ each of which consists of the superposition of two states which only differ in the phase $\Delta\phi$. In particular, the wave function of the center-of-mass motion for the ground state reads
\begin{equation*}
\ket{\psi_g}=\frac{1}{\sqrt{2}}\Big[1+e^{i\Delta\phi}\Big]\ket{\tilde\psi}\,,
\end{equation*} 
where we have introduced the abbreviation $\ket{\tilde\psi}=-\hat U_l \hat U'_l \ket{\psi}$.
The phase turns out to be half the enclosed area in phase space shown in Fig.~\ref{fig:GeneralTwoPaths}.

Needless to say, the Wigner function of $\ket{\psi_g}$ does not contain this phase. However, a phase space approach to the atomic fountain must contain this phase. It arises when we include the internal degrees of freedom into the phase space description. Unfortunately, this goes beyond the scope of the present paper.


\end{document}